# Percolative Mechanism of Aging in Zirconia-Containing Ceramics for Medical Applications

C. Pecharromán[1], J.F. Bartolomé[1], J. Requena[1], J.S. Moya[1], S. Deville[2], J. Chevalier[2], G. Fantozzi[2], R. Torrecillas[3]

[1] Instituto de Ciencia de Materiales de Madrid (ICMM), CSIC, Cantoblanco 28049, Madrid, Spain

[2] Department of Research into the Metallurgy and Physical Properties of Materials, Associate Research Unit 5510, National Institute of Applied Sciences 69621 Villeurbanne, France

[3] Instituto Nacional del Carbón CSIC, C/Francisco Pintado Fe, 26 La Corredoira, 33011 Oviedo, Spain

**Abstract**

Recently, several episodes of fracture of zirconia ceramic femoral heads of total hip prostheses have alarmed the medical and scientific community regarding aging problems in zirconia prostheses. Such fractures cause immediate local tissue reactions, which require urgent medical intervention to prevent further complications. As a result, it has been promoted that yttria-stabilized zirconia (Y-TZP) hip prostheses be substituted by alumina and alumina/Y-TZP ceramics. In the present investigation, we have found an upper limit of Y-TZP concentration in alumina/Y-TZP composites (16 vol.-%) to avoid future aging problems. This limit coincides with the percolation threshold measured by infrared (IR) reflectance in a series of alumina/Y-TZP composites.

Until now, zirconia hip prostheses were a very popular alternative to alumina implants because of their higher mechanical resistance. Y-TZP indeed exhibits the largest value of crack resistance of all the monolithic ceramics. In particular, the use of zirconia has opened the way towards new implant designs that were not possible with alumina. It has been estimated that about 500 000 patients have been implanted with zirconia femoral head prostheses since 1985, 200 000 during the past four years. Nevertheless, alarming problems related to aging have been reported recently. In particular, resistance to steam sterilization and the hydrothermal stability of yttria containing zirconia in the body have been questioned. Aging occurs by a slow tetragonal-to-monoclinic phase transformation of grains on any surface in contact with water or body fluids. [1] This transformation leads to surface roughening, [2] grain pullout, and microcracking. [3,4] Steam sterilization has been associated with surface roughening of zirconia ceramic heads. [5] As a consequence of this roughening, increase wear of the hip components can cause premature failure and requires early revision. The recognized failure rate of these products was 1 in 10 000, but nowadays, current reports indicate a failure rate as high as 8.8 % in some specific batches. [6] The U.S. Food and Drug Administration (FDA) cautions today against steam sterilization of zirconia implants. [7] More importantly, during the last year, the FDA and the Australian Therapeutic Goods Administration (TGA) announced that firms making orthopaedic implants were recalling series of Y-TZP hip prostheses due to an instance of fracture of zirconia ceramic femoral heads. [6,8] This recall follows the action by the French Agency for the Medical Safety of Health Products (AFSSAPS) and the United Kingdom Medical Devices Agency suspending the sales of Y-TZP ceramic heads. According to the manufacturer of the zirconia ceramic, the failure's origin was an accelerated tetragonal-to-monoclinic phase transformation of zirconia in particular batches. [9]

Because of these events, there is a trend today to develop advanced materials such as alumina-zirconia composites, to combine the positive effects of alumina and zirconia while avoiding the negative influence of each. In the recent literature concerning alumina-zirconia composites for biomedical applications, different compositions have been tested, from the zirconia-rich to the alumina-rich. [10-12] One route was already developed by companies (CeramTec BIOLOX delta) with the addition of 3Y-TZP (with an amount of approx. 17 vol.-%) to alumina to reach values as high as $K_{IC}$=8.5 MPa.m$^{1/2}$ and $\sigma_f$ = 1150 MPa. [13,14] However, so far, little attention has been paid to the stability of these biomedical-grade zirconia-containing materials. In this regard, it is the aim of this work to study the aging behavior of alumina/3Y-TZP composites to predict future failure of the alumina/3Y-TZP prosthesis implanted in patients by using the concept of percolation threshold.

It is now established that the tetragonal-to-monoclinic phase transformation in zirconia occurs by nucleation and growth. [2] Transformation appears first on isolated grains on the surface, then propagates to the neighboring grains as a result of various stresses and the accumulation of microcracks. The transformation penetrates step by step inside the bulk of the material, providing there is contact between zirconia grains. In the case of alumina/3Y-TZP composites, this phenomenon should require a continuous path of zirconia grains. The geometrical percolation concept, i.e., the infinite cluster formed at the percolation threshold, could therefore be applied to explain the kinetics of the aging process through the bulk of the materials, and to predict the amount of zirconia that can be added to alumina without causing aging problems.

It is well-known that experimental determination of percolation thresholds in heterogeneous composites can only be achieved when there is a large contrast between the physical properties of the different phases: [15±18] i.e., permittivity ($\varepsilon$), electrical conductivity ($\sigma$), thermal conductivity (k), etc. In the case of alumina/3Y-TZP composites, the magnitudes of these variables are roughly similar in most of the spectral regions. In addition, the strong anisotropic character of alumina (the IR zirconia dielectric tensor has been taken to be isotropic, in accordance with the literature [19]) requires that one take into consideration the contrast function between two different tensor magnitudes. As a result, it is extremely difficult to experimentally determine the critical concentration. There is one exception: it has be found that around the infrared region of the longitudinal phonons of alumina (from 900 cm$^{-1}$ to 1 000 cm$^{-1}$), the complex dielectric tensor of both alumina and zirconia [19,20] are nearly isotropic and display a significant contrast (Fig. 1), which can be as high as $\|\varepsilon_{alumina}\| \leq 100\|\varepsilon_{zirconia}\|$.

In order to simplify the analysis of the IR-effective properties in the considered region of the spectrum, we will assume an isotropic model for the IR dielectric constant of Al$_2$O$_3$:

$$\varepsilon(\omega) = \varepsilon_\infty \frac{(\omega_L^2 - \omega^2) - i\gamma\omega}{(\omega_T^2 - \omega^2) - i\gamma\omega} \qquad \text{(eq. 1)}$$

This approximation can be done only in this spectral region, owing to the fact that the highest transversal and longitudinal phonons of alumina for both A2u (589 cm$^{-1}$ and 871 cm$^{-1}$) and Eu (635 cm$^{-1}$ and 900 cm$^{-1}$) modes are very close. Then, the transversal and longitudinal frequencies, $\omega_T$ and $\omega_L$, respectively, the damping $\gamma$, and the optical dielectric constant $\varepsilon_\infty$ can be approximately taken as the weighted average of both orientations ($\omega_T$=620 cm$^{-1}$, $\omega_L$=890 cm$^{-1}$, $\gamma$=15.1 cm$^{-1}$, and $\varepsilon_\infty$=3.2). Although the permittivity cannot be measured directly with conventional IR spectrometers, the specular reflectance is a function of $\varepsilon$ and must reveal any large changes in this function. The near-normal specular reflectance is given by:

$$R = \left|\frac{\sqrt{\varepsilon}-1}{\sqrt{\varepsilon}+1}\right|^2 \quad \text{(eq. 2)}$$

The combination of Equation 1 for the dielectric constant with Equation 2 means that the reflectance curve has a broad band with values for R very close to 1, which extends from $\omega_T$ to $\omega_L$ and quickly descends for frequencies larger than $\omega_L$. In fact, R reaches a minimum at the approximate frequency which satisfies:

$$\sqrt{\varepsilon(\omega_m)} - 1 \cong 0 \quad \text{(eq. 3)}$$

This frequency is approximately $\omega_m$=1 000 cm$^{-1}$ for pure polycrystalline alumina. Therefore, the range of reflectance minimum of alumina/3Y-TZP composites studied extends from the longitudinal modes of alumina ($\omega_L$=900 cm$^{-1}$) to $\omega_m$=1 000 cm$^{-1}$ (Figs. 1, 2a).

The reflectance profile of those composites depends on the shape of the effective dielectric constant below and above the percolation threshold at the frequencies $\omega_L$ and $\omega_m$, respectively. For 3Y-TZP concentrations below the percolation threshold ($f<f_c$), the effect of the addition of zirconia to the composite on the effective dielectric constant for the whole spectral range is a small increase. This implies that the condition of Equation 3 is satisfied for lower frequencies (Fig. 1). It results in a reflectance profile similar to that of pure alumina with the reflectance minimum shifted to slightly lower frequencies. However, above the percolation threshold, the effective dielectric constant will rise sharply at $\omega = \omega_L$, where the dielectric contrast is a maximum, and moderately at $\omega = \omega_m$, where the dielectric contrast is less than 4. This occurs in such a way that the effective dielectric constant for the whole spectral region from $\omega_L$ to $\omega_m$ will take values similar to that of $\varepsilon_\infty$ of zirconia ($\varepsilon_\infty = 4.2$). Such region of small dispersion of $\varepsilon(\omega)$ will also induce a nearly flat area to occur in the reflectance spectrum.

This behavior is clearly detected experimentally. As it can be seen in Figure 2a, the minimum of the reflectance curve gradually shifts to lower frequencies, for concentrations lower than f = 0.14. Above this point, the shape of the minimum located at 1 000 cm$^{-1}$ drastically changes. In Figure 2b, this effect is represented as a plot of the reflectance minimum vs f, then it is clear that for f > 0.14 the reflectance increases dramatically. This fact suggests that the percolation threshold of alumina/3Y-TZP composites must be located around $f_c \approx 01.6$ as predicted using standard models for three-dimensional percolation. [15,16]

Figure 3 shows the variation in the fraction of monoclinic phase vs time of steam sterilization for some selected compositions. Monoclinic content was measured by an X-ray diffraction (XRD) technique (Cu-K$\alpha$ radiation). When values of mass absorption coefficients and densities of the studied alumina/$ZrO_2$ composites are taken into consideration, [21] this technique gives information on penetration depths (defined as an attenuation of $1/e = 0.37$ for a beam normal to the surface) ranging from 50 µm to 15 µm, depending on the fraction of zirconia. The penetration depth decreases linearly with an increasing vol.-% of $ZrO_2$. Therefore, when there are more than 10 consecutive layers of grains, we can consider that these data supply an accurate representation of processes happening in three dimensions. For all the materials studied, a first increase of up to 10 % of monoclinic phase was observed after two hours. This first increase was easily related to the presence of some (scarce) 3Y-TZP aggregates in the materials (Fig. 4a). More importantly is that for specimens with zirconia content of less than or equal to 14 vol.-%, after this initial slight increase, no more evolution was observed, even after 100 hours in the autoclave. On the other hand, when $f>f_c$, the monoclinic phase fraction continues to grow without reaching saturation. This clearly means that transformation occurs by a progressive invasion of the material (Fig. 4b). Figure 5 represents the amount of transformation observed using XRD analysis after 40 hours of aging in steam, as compared with the initial 3Y-TZP content. This correlates especially well with the percolation theory. Above the critical amount of 16 vol.-% 3Y-TZP, geometrical percolation allows a continuous path between zirconia particles, so that transformation can proceed.

One might question the correlation of these accelerated tests with the reality of the in-vivo situation. The activation energy of zirconia aging [2] is of the order of 100 kJ mol$^{-1}$. These accelerated tests in steam give a rough estimation of the monoclinic fraction obtained after a hundred years at 37°C. However, it is also very often stated that during wear, the temperature of the surface of the prosthesis can reach as high as 50°C. [22] This temperature would give a more pessimistic estimation of only several years. In any case, taking into consideration the uncertainties in the real temperature, in the estimation on the activation energy, and the recent events of unexpected accelerated aging of zirconia, no aging at all should be acceptable.

In summary, it has been shown that comparing specular IR reflectance measurements with aging experiments, the concept of the percolation threshold is relevant when talking about degradation resulting from aging. As far as biomedical applications are concerned, zirconia-toughened alumina ceramics would be very appropriate materials, provided that the zirconia content is kept below the percolation threshold. This study has established an upper limit of 16 vol.-% 3Y-TZP inside an alumina matrix. Therefore, the authors caution ceramics manufacturers and surgeons against fabrication of zirconia-alumina composites, where the amount of zirconia is above 16 vol.-% (22 wt.-%).

## Experimental

Plates with 80x80x3 mm³ dimensions ranging from 7 vol.-% 3Y-TZP in an alumina matrix to pure 3Y-TZP were obtained by slip casting in a plaster of Paris mold and sintered to >99 % density at 1600C for 2 h in the case of alumina-zirconia composites, and at 1450°C for 2 h in the case of pure 3Y-TZP. They were processed from high-purity biomedical-grade powders (Tosoh TZ3Y and Condea HPA 0.5). Samples were polished with diamond paste in order to reach the surface topography recommended for hip prostheses. The initial monoclinic content was in all cases close to zero. Aging experiments were carried out by leaving samples in a steam autoclave at a temperature of 140°C. The diamond-polished side of each specimen was examined by X-ray diffraction (XRD) before and after aging. XRD data were obtained with a diffractometer using Ni-filtered Cu-Kα radiation. The tetragonal/monoclinic zirconia ratio was determined using the integrated intensity (measuring the area under the diffractometer peaks) of the tetragonal (101) and two monoclinic (111) and (1-11) peaks as described by Garvie et Nicholson [23] and then revised by Toroya et al. [24,25]. For the purpose of comparison, the integrated intensities obtained were individually normalized to the (101) tetragonal integrated intensity for each composition. Diffractometer scans were obtained from 2θ = 27-33° at a scan speed of 0.2°min$^{-1}$ and a step size of 0.02°. Near-normal specular reflectance spectra were taken on diamond-polished surfaces of the different alumina/3Y-TZP composites with a near-normal reflectance attachment (12°) in a FTIR Bruker 66V spectrometer.

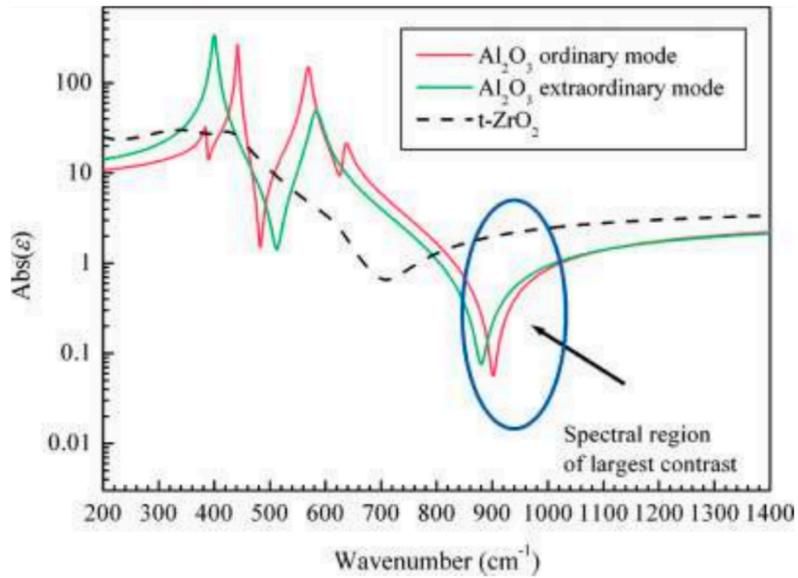

Fig. 1. Absolute value of the dielectric constant for the ordinary (Eu) and extraordinary (A2u) modes of alumina and tetragonal zirconia.

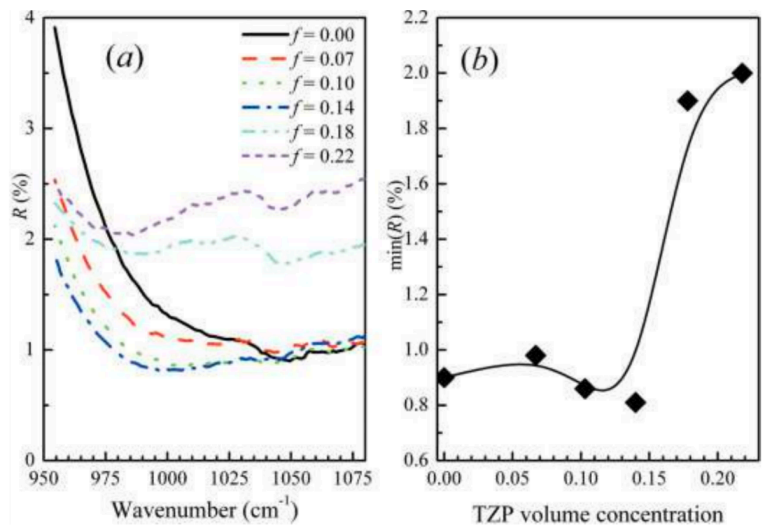

Fig. 2. a) Near-normal reflectance spectra of alumina/ 3Y-TZP composites in the region of the minimum of reflectance. b) Minimum of reflectance vs 3Y-TZP volume fraction (f). The solid lines are guides to the eyes.

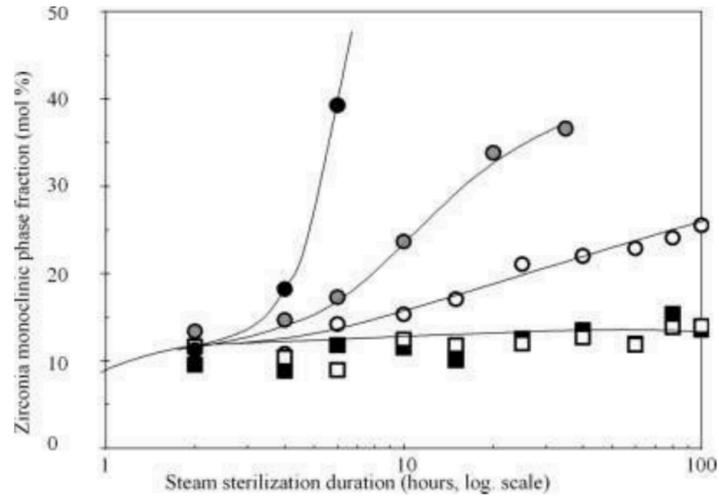

Fig. 3. Fraction of monoclinic phase vs time of steam sterilization for alumina/3Y-TZP composites with different zirconia content. □: 7 vol.-% zirconia, ■: 14 vol.-%, ◖: 22 vol.-%, : ○ 40 vol.-%, d: ● 100 vol.-%.

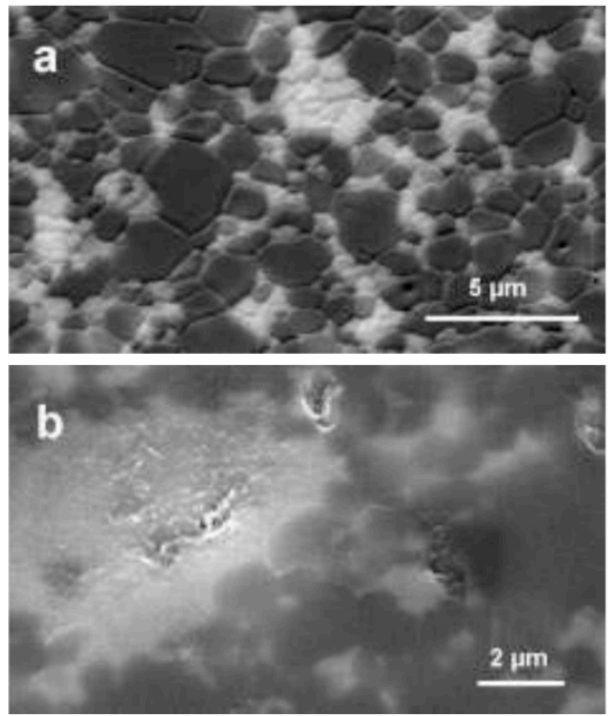

Fig. 4. SEM micrographs of polished surface of alumina/22 vol.-% 3Y-TZP a) before and b) after 115 h steam sterilization.

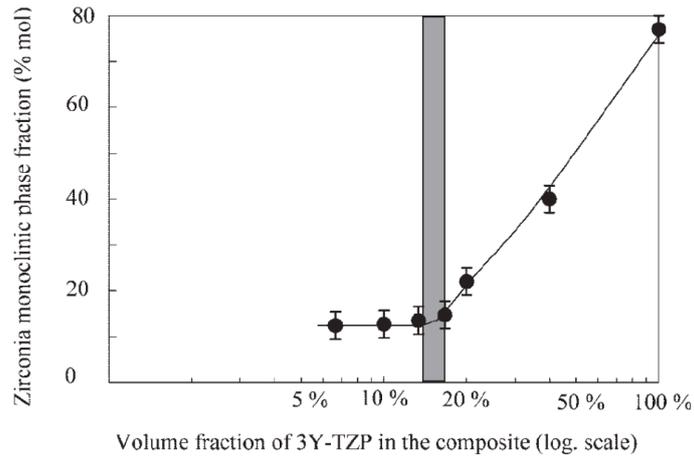

Fig. 5. Fraction of monoclinic phase measured after 40 h steam sterilization as a function of the zirconia volume content in the composite. The dashed area represents the percolation threshold measured by IR study.